\documentclass[twocolumn,pra,superscriptaddress,noshowpacs]{revtex4}

\usepackage{graphicx,nicefrac}
\usepackage{amssymb}
\usepackage{amsfonts}
\usepackage{xcolor}
\usepackage{amsmath}
\usepackage{siunitx}
\usepackage{soul}

\usepackage[utf8]{inputenc}

\usepackage{amsmath}
\usepackage{mathtools}
\usepackage{braket}
\usepackage{dsfont}
\usepackage{bigints}
\usepackage{graphicx}
\usepackage{amssymb}
\usepackage{xcolor}
\usepackage{bbold}

\graphicspath{ {./Images/} }

\usepackage{adjustbox}
\makeatletter
\newcommand{\fitimage}[2][\@nil]{
	\begin{figure}
		\begin{adjustbox}{width=0.9\textwidth, totalheight=\textheight-2\baselineskip-2\baselineskip,keepaspectratio}
			\includegraphics{#2}
		\end{adjustbox}
		\def\tmp{#1}%
		\ifx\tmp\@nnil
		\else
		\caption{#1}
		\fi
	\end{figure}
}
\makeatother

\newcommand{\approxpropto}{\mathrel{\vcenter{
			\offinterlineskip\halign{\hfil$##$\cr
				\propto\cr\noalign{\kern2pt}\sim\cr\noalign{\kern-2pt}}}}}

\begin{document}

\title{Linear and nonlinear edge dynamics of trapped fractional quantum Hall droplets}
\author{Alberto Nardin}\email{alberto.nardin@unitn.it}
\author{Iacopo Carusotto}
\affiliation{INO-CNR BEC Center and Dipartimento di Fisica, Universit\`a di Trento, via Sommarive 14, I-38123 Trento, Italy.}

\begin{abstract}
We report numerical studies of the linear and nonlinear edge dynamics of a non-harmonically confined macroscopic fractional quantum Hall fluid. 
In the long-wavelength and weak excitation limit, observable consequences of the fractional transverse conductivity are recovered.
The first non-universal corrections to the chiral Luttinger liquid theory are then characterized: for a weak excitation in the linear response regime, cubic corrections to the linear wave dispersion and a broadening of the dynamical structure factor of the edge excitations are identified; for stronger excitations, sizable nonlinear effects are found in the dynamics.
The numerically observed features are quantitatively captured by a nonlinear chiral Luttinger liquid quantum Hamiltonian that reduces to a driven Korteweg-de Vries equation in the semiclassical limit. Experimental observability of our predictions is finally discussed.
\end{abstract}

\maketitle

\section{Introduction}
The fractional quantum Hall (FQH) effect is one of the most fascinating concepts of modern quantum condensed matter physics~\cite{Jain,Tong}. Whereas FQH states of matter were originally observed in the solid-state context of two-dimensional electron gases under strong magnetic fields, a strong experimental attention is presently devoted to synthetic quantum matter systems~\cite{HannahTomoki} such as gases of ultracold atoms under synthetic magnetic fields~\cite{ReviewNigel,review_Goldman_synthetic,review_dalibard,TopoBandsForAtoms} or fluids of strongly interacting photons in nonlinear topological photonics devices~\cite{RMPQFL,Topo,NPHYS}. As it was pointed out in recent theoretical proposals~\cite{Paredes,Cooper2,Regnault,Rizzi,Rougerie,Morris,QHWithOptFluxes, Goldman1,Goldman,Unal,Onur,Onur2,Onur3,Onur4,AMDLH}, such systems typically offer a wider variety of experimental tools compared to the transport and optical probes of electronic systems. Important experimental steps towards observing FQH physics have been recently reported in both atomic~\cite{Gemelke,Greiner,Leonard} and photonic systems~\cite{Roushan,Clark}.

One of the most exciting features of FQH liquids is the possibility of observing fractional statistics effects both in the bulk and on the edge~\cite{Stern,Nayak}. On this latter, in particular, gapless modes supporting fractionally charged excitations have been observed in shot-noise experiments~\cite{DePicciotto}; more recently, edge modes have been used as a probe of the topological state of the bulk~\cite{Nakamura}, hints of generalized exclusion statistics have been highlighted~\cite{DeBartolomei},
and  a number of further intriguing properties have been anticipated~\cite{FurtherIntriguingProperties,FurtherIntriguingProperties2}. Many of these features are theoretically captured by the chiral Luttinger liquid ($\chi$LL) theory~\cite{Wen_topo, Wen2, Chang} which is expected to be an accurate description of the edge in the long-wavelength and weak excitation limits.

In this work, we investigate the physics  beyond the the regime of validity of the $\chi$LL description and perform numerical studies of the linear and nonlinear edge dynamics of a {\em fractional} QH liquid trapped by a generic, non-harmonic external potential. 
As compared to our previous study of  {\em integer} QH liquids~\cite{ourepl}, the strongly correlated nature of FQH liquids poses enormous technical challenges to the theoretical description 
and requires the development of a novel numerical approach to follow the dynamics of macroscopic FQH clouds.
In particular, we focus on the neutral edge excitations (EE) 
that are generated by applying an external time-dependent potential to an incompressible FQH cloud.

In electronic systems generation and diagnostics of edge excitations requires ultrafast tools that are presently being developed with state-of-the-art electronic and optical technologies~\cite{Yusa,Ashoori}. On the other hand, arbitrary time-dependent potentials can be readily applied to synthetic systems and high-resolution detection tools at the single-particle level are also available~\cite{HannahTomoki}. This suggests that our results will offer a useful guidance to the next generation of FQH experiments in a wide range of experimental platforms. 

In addition to this, we expect that our results may be of interest also from a theoretical perspective: leveraging on the physical insight provided by numerical calculations, we are able to formulate a nonlinear extension of $\chi$LL theory that is able to quantitatively describe the system dynamics at a much lower numerical cost. This theory offers an effective theoretical framework for future investigations of the rich nonlinear quantum dynamics of the FQH edge and is amenable to sophisticated theoretical tools for non-linear Luttinger liquids~\cite{Glazman}.


The structure of the article is the following.
In section \ref{section:II} we discuss the physical system under consideration (\ref{subsection:II.A}), we introduce our numerical approach for its description (\ref{subsection:II.B}), and we show some benchmark calculations (\ref{subsection:II.C}).
In section \ref{section:III}, signatures of the quantized transverse conductivity of the bulk in the edge physics are highlighted and discussed within the $\chi$LL picture. 
In section \ref{section:IV} we start investigating effects beyond the $\chi$LL description by looking at the dynamical structure factor of a anharmonically confined droplet  (\ref{subsection:IV.A}), at the group velocity dispersion of the edge excitations (\ref{subsection:IV.B}) and at their nonlinear features at stronger excitation levels (\ref{subsection:IV.C}).
In section \ref{section:V}, we capitalize on the numerical observations of the previous section to write a minimal non-linear $\chi$LL Hamiltonian whose classical limit gives a Kortweg-de Vries equation for the edge-density dynamics. In particular, we show how this generalized $\chi$LL Hamiltonian is able to reproduce all the microscopic calculations in a quantitative way. In section \ref{section:VI} we discuss the experimental observability of the described physics.
Finally, we give some conclusive remarks in section \ref{section:VII}.
The Appendices summarize additional information in support of our claims: Appendix \ref{appendix:A} shows statistical information on the collected Monte Carlo data; 
in appendix \ref{appendix:B} we comment on the protocol we used to excite the edge dynamics;
Appendix \ref{appendix:C} provides further details on the linear response calculations within the $\chi$LL theory;
in Appendix \ref{appendix:D} we show some additional data on the broadening of the dynamic structure factor due to the anharmonic confinement; in Appendix \ref{appendix:E} we compare the microscopic numerical results  for the time evolution with the nonlinear $\chi$LL description.

\section{The physical system and the numerical method}
\label{section:II}

    \subsection{The physical system}
    \label{subsection:II.A}
    We consider a 2D system of $N$ quantum particles 
    with short-range repulsive interactions subject to 
    a uniform magnetic field $B$ orthogonal to the plane. 
    In this continuous-space geometry with no underlying periodic lattice, the single-particle states in a uniform $B$ organize in highly degenerate and uniformly separated Landau levels: in what follows, energies are measured in units of the cyclotron splitting between Landau levels and lengths  in units of the magnetic length, with the usual complex-valued shorthand $z=x+iy$.
    Two-body interactions lift the degeneracy and lead to the formation of highly-correlated incompressible ground states. The simplest examples are the celebrated Laughlin states (LS)~\cite{laughlin,Wavefunctionology}
    \begin{equation}
        \label{eq_Laughlin}
       \Psi_L(\left\{z_i\right\})=\prod_{i<j}(z_i-z_j)^{1/\nu}\,\exp\left(-\sum_{i}\left|z_i\right|^2/4\right),
    \end{equation}
 entirely sitting within the lowest Landau level (LLL).
    The LS at filling $\nu=1/2$ is the exact ground state for contact-interacting bosons~\cite{wilkingunn,elia}; the $\nu \neq 1/2$ LS is the exact GS of certain bosonic or fermionic toy model Hamiltonians~\cite{HaldanePseudoPotentials, Trugman, SimonRezayiCooper} and an excellent approximation in more realistic cases.
    
    In this work, we focus our attention on the gapless EE on top of a LS. These excitations correspond to chirally-propagating surface deformations of the incompressible cloud and, in the low-energy/long-wavelength limit, are accurately described by the $\chi$LL model~\cite{Wen_topo, Wen2, Chang, Caz}. 
    Our goal is to understand the basic features of the dynamics beyond the $\chi$LL description, when the cloud is confined by a generic non-harmonic trap potential $V_\text{conf}(r)=\lambda r^\delta$ and the applied time-dependent 
    excitation strength is large enough to exit the linear regime. 
    To keep the calculation manageable, we will assume that the trap is shallow enough and the excitation is not too strong, so to avoid coupling to states above the many-body energy gap $\Delta$~\cite{Regnault, RegnaultJolicoeur, ReviewNigel}. In this way the ground state remains a Laughlin state and the dynamics of the system edge is confined to the subspace of many-body wavefunctions obtained by multiplying the Laughlin wavefunction by holomorphic symmetric polynomials $P_{\alpha}\left(\left\{z_i\right\}\right)$ of the particle coordinates~\cite{Stone,Caz,Wen_topo,Wavefunctionology}.
    
    In order to make these qualitative considerations more precise and quantitative, we can note that the Laughlin state remains the GS in the presence of the confinement potential as long as the energy cost of adding a particle at the system edge is smaller than the one required for inserting the extra particle into the bulk of the system, that is the many-body gap. 
    Under this assumption, the GS is a everywhere Laughlin state and its edge states are well captured by our theory~\footnote{This of course holds unless the edge is reconstructed\cite{ChamonWen}, which however does not typically happen for local interactions.}.
    If the aforementioned condition is not strictly met, a shell structure of locally homogeneous incompressible liquids has been predicted to appear~\cite{cooper_wedding_cakes}, separated by sudden jumps at the transition points between different strongly correlated liquids at different filling fractions. In spite of this additional complication, we expect that our theory will still provide an accurate description at least of the external edge between the outer Laughlin shell and the external vacuum, provided the outer shell is thicker than the characteristic correlation length of the gapped bulk, of the order of the magnetic length.
    
While we expect that our results can be generally applied to a variety of systems in different geometries, it is interesting to have a closer look at the relevant energy scales for the promising case  of rotating clouds of bosonic atoms~\cite{ReviewNigel,Gemelke,Zwierlein2}: atoms are confined to move along a two-dimensional plane by a tight confinement along $z$ and are laterally trapped by a harmonic $V_2=\frac{1}{2}M\omega^2r^2$ potential supplemented by a anharmonic $V_{\rm conf}=\lambda r^\delta$ one. In the fast rotation regime at {$\Omega_r=\omega$}, the centrifugal potential in the rotating frame is completely compensated by the harmonic part of the confinement and one is left with the anharmonic trapping only.
    Given the tight confinement along $z$, the effective two-body interaction potential is a contact one, {$V_\text{int}=2\hbar\Omega_r g \delta^{(2)}(\mathbf{r}/l_B^2)$} with an interaction strength $g$ proportional to the ratio $a_S/a_z$ between the s-wave scattering length and the harmonic oscillator length $a_z$. As usual, {$l_B=\sqrt{\hbar/(2M\Omega_r)}$} is the effective magnetic length and {$2\hbar\Omega_r$} the effective cyclotron gap.
    
    {The characteristic energy scale of the interactions between the bosons is thus $V_\text{int}=2\hbar\Omega_r g (n_{2D}l_B^2)$, where $n_{2D}$ is the two-dimensional density of the gas:
    for a Laughlin state at half filling $n_{2D}=1/(4\pi\,l_B^2)$, so the interaction energy scale is of order $V_\text{int}/2\hbar\Omega_r=g/4\pi$. 
    While the quantum correlations between particles make the Laughlin state and its edge excitations exact zero-energy eigenstates of the Hamiltonian, the typical energy of quasi-particle excited states is set by $V_\text{int}$. In particular, numerical calculations~\cite{ReviewNigel} have shown that the many-body energy gap in these systems is of the order of $\Delta\approx0.1\,g\,\hbar\Omega_r$.}
    While the dimensionless parameter $g$ could be tuned to relatively large values by means of Feshbach resonances, in our case it is beneficial to keep it moderate $g/(4\pi)\lesssim1$ so as to suppress the Landau level mixing. As a result, one can expect optimal values of the many-body gap to be on the order of a fraction of {$\hbar\Omega_r$}, which visibly points in the direction of using strong in-plane harmonic potentials.
    
As a final point, we can spell out the condition not to break the many-body gap in the case of a quartic $\delta=4$ anharmonic confinement potential. 
This requires that at the position of the edge ($r\sim\sqrt{2N/\nu}\,l_B$) the anharmonic part is much smaller than the harmonic one, $V_{\rm conf}/V_2 \sim 10^{-3}$. 
    For a system of $N=25$ particles, for which as we will show in what follows the physics already approaches the thermodynamic limit, this condition sets the magnitude of the anharmonic potential to be roughly {$V_{\rm conf}(R_\text{cl})/2\hbar\Omega_r\approx0.01$}, which imposes {$\hbar\lambda/M^2\Omega_r^3\approx 10^{-5}$}. Under these conditions the ground state of the system will be the bosonic Laughlin state at half filling.
    Scaling up the size of the system will in principle require precise control on the trap parameters; however, as we briefly discuss in Sec.\ref{section:VI}, we expect the same physics to emerge for a general confining potential.
    
    \subsection{The numerical approach}
    \label{subsection:II.B}
    
    We expand the many-body wavefunction $\Psi$ over these many-body states as
    \begin{equation}
        \Psi(\left\{z_i\right\})=\sum C_{\alpha} P_{\alpha}(\left\{z_i\right\}) \,\Psi_{L}(\left\{z_i\right\})\,,
        \label{eq:Laughlin+poly}
    \end{equation}
    where 
    $\alpha$ runs through the angular momentum $l$ sectors and through the $p_N(l)$ states corresponding to the integer partitions of $l$ restricted to $N$ elements at most, which span each $l$ sector.  
    Projecting the many-body Schr\"odinger equation over these basis states, we obtain a Schr\"odinger equation
    \begin{equation}
        \label{eq_projectedSchrodinger}
        i\,\mathds{M}_{\beta,\alpha} \dot{C}_{\alpha}= \mathds{H}_{\beta,\alpha}C_{\alpha}
        \end{equation}
        for the expansion coefficients $C_\alpha$.
    
    The kinetic energy is constant within the LLL and the two-body interaction energy is assumed to be negligible within the subspace of Laughlin-like states (it is exactly zero for the case of contact-interacting bosons). The Hamiltonian $\mathds{H}$ then only includes the confinement potential $V_{\rm conf}(r)$, and the ``metric'' $\mathds{M}$ accounts for the non-orthonormality of the basis wavefunctions.
    A similar approach was previously adopted to study the ground-state properties and the spectrum of EE of a FQH fluid of Coulomb-interacting fermions~\cite{MC_Jain0, MC_Jain1, MC_Jain2, MC_Jain3,MC_Jain4}; here we make a crucial step forward and apply it to the time-dependent dynamics of the strongly correlated FQH fluid, in particular to its response to an external potential $U$. 
    
    The great advantage of our approach is that it allows to tame the dimension of the many-body Hilbert space: for a given $l$, the dimension of the Hilbert subspace does not grow with $N$. 
    The price is the need to compute the high-dimensional integrals hidden in the matrix elements of $\mathds{H}$ and $\mathds{M}$: in our calculations, this is done by means of a Monte Carlo sampling of the many-body wavefunction via a standard Metropolis-Hastings algorithm with a weight that generalizes to excited states the well-known Laughlin's plasma analogy~\cite{Tong}.
    
    Specifically, the calculation of the matrices $\mathds{M}$ and $\mathds{H}$ appearing in \eqref{eq_projectedSchrodinger} require the evaluation of matrix elements of a generic real-space observables $\mathcal{O}(\{z_i\})$ between two (non-necessarily normalized) many-body states $\psi_{1,2}(\{z_i\})$. This quantity can be rewritten as:
    \begin{multline}
        \int\,\mathcal{D}z\,\frac{\psi_1^*(z)}{\sqrt{\| \psi_1 \|^2}} \mathcal{O}(z) \frac{\psi_2(z)}{\sqrt{\| \psi_2 \|^2}}=\\=\frac{
        \int\,\mathcal{D}z\,\frac{|\psi_1(z)|^2}{\|\psi_1\|^2}\,\frac{\mathcal{O}(z)\,\psi_2(z)}{\psi_1(z)}}{\sqrt{ \int\,\mathcal{D}z\,\frac{|\psi_1(z)|^2}{\|\psi_1\|^2} \left| \frac{\psi_2(z)}{\psi_1(z)}\right|^2}}
    \end{multline}
    where we have introduced the short-hands $z=\{z_1\ldots z_N\}$ and $\mathcal{D}z=dz_1\ldots dz_N$ and we have defined the norm as $\|\psi_{1,2}\|^2=\int\,\mathcal{D}z\,|\psi_{1,2}(z)|^2$. The integrals in both the numerator and the denominator are then performed with the Metropolis-Hastings algorithm using $W(z)={|\psi_1(z)|^2}/{\|\psi_1\|^2}$ as the target probability distribution function~\cite{metropolis,hastings}. Since the $\psi_{1,2}(z)$ wavefunctions have the form \eqref{eq:Laughlin+poly} consisting of a Laughlin state multiplied by a suitable polynomial of moderate degree, they share most of their zeros and their weights are concentrated in similar regions of configuration space. This feature is strongly beneficial in view of the convergence of the Monte-Carlo sampling. In principle, the matrices $\mathds{M}$ and $\mathds{H}$ obtained in this way are not exactly Hermitian, so we perform a preliminary Hermitization step before proceeding with the calculations.

    Using this method we have been able to study the dynamics of systems of up to $N\sim 80$ particles. In the following we will focus on results for up to $40$ particles for which the statistical error of the Monte Carlo sampling is smaller (see Appendix \ref{appendix:A}). As we are going to see, for this particle number, the system is in fact large enough to be in the macroscopic limit where the edge properties are independent of the system size.
  
        \begin{figure}[h]
    	\begin{adjustbox}{width=0.48\textwidth, totalheight=\textheight-2\baselineskip,keepaspectratio,right}
        	\includegraphics[]{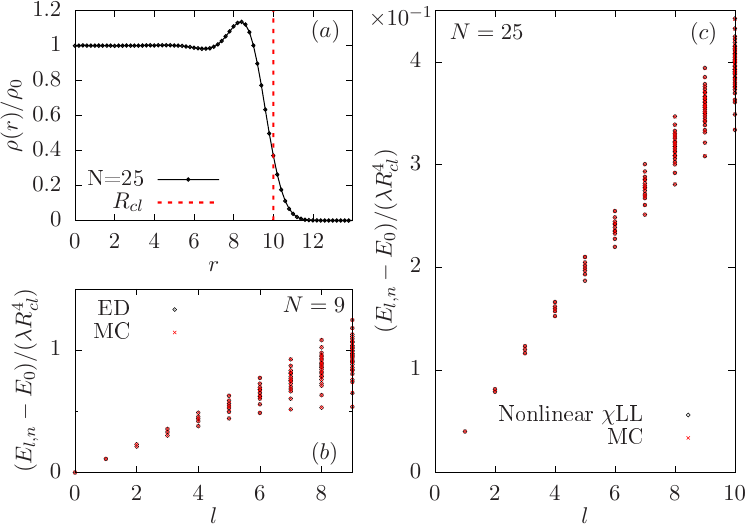}
        \end{adjustbox}
        \caption{
        (a) Radial profile of the GS density.
        (b,c) Excitation spectra for (b) $N=9$ and (c) $N=25$ particles (red crosses), compared to ED [black dots in (b)] and the nonlinear $\chi$LL theory \eqref{eq_NonLinearXLL} [black dots in (c)].
        Anharmonic $\delta=4$ trap with $\lambda=10^{-6}$, filling factor $\nu=1/2$.
        \label{fig_GSDensity_Spectrum}}
    \end{figure} 
    
     \subsection{Benchmark}
     \label{subsection:II.C}
    A first application of the numerical MC method is illustrated in Fig.\ref{fig_GSDensity_Spectrum} where we show a radial cut of the GS density (a) and the energies of the lowest-$l$ excited states sitting below the many-body energy gap (b,c). 
    The density profile shows the density plateau corresponding to the incompressible bulk $\rho_0=\nu/\left(2\pi\right)$ and the usual oscillating structure on the edge near the classical radius $R_{cl}=\sqrt{2 N/\nu}$~\cite{Tong}. 
    The excited state energies successfully compare to exact diagonalization (ED) results for all particle numbers for which ED is feasible (b). 
    
        \begin{figure}[h]
        \begin{adjustbox}{width=0.48\textwidth, totalheight=\textheight-2\baselineskip,keepaspectratio,right}
	        \includegraphics[]{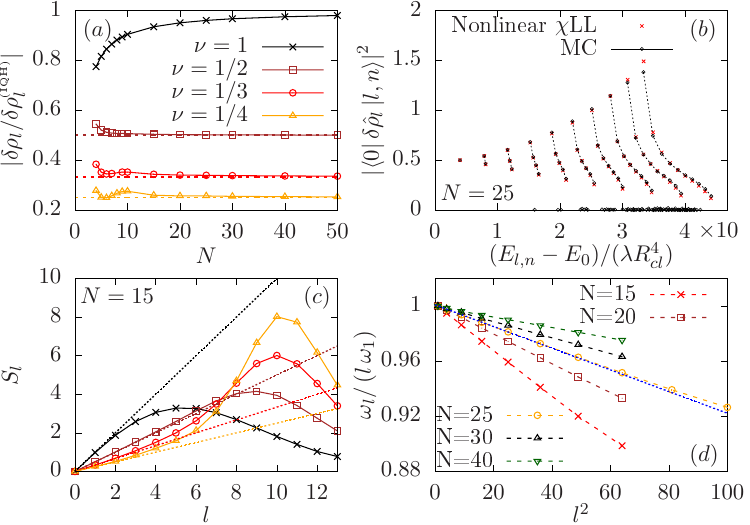}
        \end{adjustbox}
        \caption{
        (a) Amplitude of the edge density response after the weak $l=2$ external potential has been switched off, for different filling factors $\nu$, normalized to the one of a large IQH system.
        (b) DSF weights plotted against the excitation energy of each eigenstate. Within each $l$ sector, the dashed lines are guides to the eye. MC data (black dots) are compared to the nonlinear $\chi$LL theory (red crosses).
        (c) SSF $S_l$ as a function of $l$ for the same values of $\nu$ as in (a). Dashed lines indicate the $\chi$LL prediction $S_l=\nu l$.
        (d) Normalized edge-mode dispersion for different $N$. Same trap potential as in Fig.\ref{fig_GSDensity_Spectrum}. In panels (b,d) the filling factor is fixed to $\nu=1/2$.
        \label{fig_bulkCurrent_SF_DSF_FreqShift}
        }
    \end{figure}   
     
 \section{Quantized transverse conductivity}
 \label{section:III}

    We then investigate the dynamical evolution of the system in response to a temporally short excitation.
    With no loss of generality 
    we assume for simplicity a radially flat potential 
    carrying definite angular momentum $l$. In the usual $(r,\theta)$ polar coordinates, the potential has the simple form
    \begin{equation}
    U(r,\theta,t)=U(\theta,t)=U_l(t)\,e^{i l\, \theta} + \textrm{c.c.}.
    \label{eq:Ul}
    \end{equation} 
    where $U_l(t)$ is the (complex-valued) time-dependent amplitude of the excitation at angular momentum $l$. Here, $l$ plays the role of a proxy of the excitation wavevector: for a fixed cloud size, the higher $l$, the shorter the effective wavelength of the excitation along the edge. While the calculations reported in the main text refer to this $r$-independent potential, the general case of a $r$-dependent $U(r,\theta,t)$ is discussed in Appendix \ref{appendix:B} and shown to bring no additional physics.
From the temporal point of view, we focus on the case of a pulsed excitation with a Gaussian temporal shape $U_l(t)=u_0\,\exp(-(t/\tau)^2)$. The characteristic time $\tau$ for turn-on and then switch-off is taken to be slow enough $\tau\gg \hbar / \Delta$ to avoid a significant excitation of states above the many-body gap, but fast enough compared to the edge mode frequencies so to induce a significant excitation of them.
    
    As expected on physical grounds, the force along the azimuthal direction induced by the angular gradient of $U(\theta,t)$ generates a transverse Hall current along the radial direction, which locally changes the cloud density on the edge.
    Numerical results for the linear response to a weak excitation are displayed in Fig. \ref{fig_bulkCurrent_SF_DSF_FreqShift}(a): in agreement with transverse conductivity quantization arguments, a clear proportionality of the response on the FQH filling factor $\nu$ is found in the large-$N$ limit. Quite remarkably, this limiting behaviour is accurately approached in the FQH case already for way lower particle numbers $N\gtrsim 15$ in the FQH than in the $\nu=1$ IQH case. This conclusion is of great experimental interest as it suggests that evidence  of the quantized conductivity can be observed just by probing the response of the edge of relatively small clouds to trap deformations, a technique of widespread use for ultracold atomic clouds~\cite{SandroBook}.

    This behaviour can be understood on the basis of the $\chi$LL theory~\cite{Wen_topo, Wen2, Chang, Caz}, with the external potential $U(\theta,t)$ minimally coupled to the edge density $\hat{\rho}(\theta)$. 
    The system response after $U$ has been turned off can be written (\ref{appendix:C}) to linear order as \begin{equation}
        \braket{\delta\hat{\rho}(\theta,t)}=\frac{1}{\pi}\Im\left[
    \sum_l
    \int \widetilde{U}_{l}(\omega) S_l(\omega)  e^{i (l \theta-\omega t)}\,d\omega \right]\,,
    \end{equation}
    where $\widetilde{U}_{l}(\omega)$ is the space-time Fourier transform of $U(\theta,t)$,
    \begin{equation}
        S_l(\omega)=\int \frac{dt}{2\pi}\,e^{i\omega t}\braket{e^{i\hat{H}t} \delta\hat{\rho}_{l} e^{-i\hat{H}t} \delta\hat{\rho}_{-l}}
    \end{equation}
    is the dynamical structure factor (DSF) --restricted here to the edge mode manifold of states-- and
    $\delta\hat\rho_l$ is the angular Fourier transform of the edge-density variation $\delta\hat{\rho}(\theta)$.
    When the trap is quadratic, the edge is a prototypical $\chi$LL and the DSF is a $\delta$-peak centered at $\omega_l = \Omega\,l$, with $\Omega=2\lambda$.
    For anharmonic traps [Fig.\ref{fig_scalings}(a)], $\Omega$ is still determined by the potential gradient at the cloud edge,
    \begin{equation}
        \label{eq:velocity}
        \Omega=\left.{r}^{-1}{\partial_r V_\text{conf}(r)}\right|_{R_{cl}}\propto N^{(\delta-2)/2}
    \end{equation}
     but at the same time the DSF broadens. 
    Up to not-too-late times, the density response can nevertheless be accurately approximated as 
    \begin{equation}
        \label{eq:edge_density_variation}
        \braket{\delta\hat{\rho}(\theta,t)}\simeq\frac{1}{\pi}\Im\left[
    \sum_l \widetilde{U}_{l}(\omega_l)\, e^{i \left(l \theta -\omega_l t\right)} S_l\right]\,,
    \end{equation}
    where $S_l=\int S_l(\omega) \,d\omega$ is the edge-mode static structure factor (SSF). 
    As long as the confinement potential is not strong enough to mix with states above the many-body gap, the SSF keeps its $\chi$LL value $S_{l}=\nu l$ for $l\geq 0$ and zero otherwise up to $l$ values where finite-$N$ effects get important [Fig.\ref{fig_bulkCurrent_SF_DSF_FreqShift}(c)].
    
\section{Beyond chiral Luttinger liquid effects}
 \label{section:IV}
Our numerical framework is not restricted to study the response of the system to weak and long-wavelength excitations as captured by the standard chiral Luttinger liquid theory. The goal of this Section is to explore the physics beyond the $\chi$LL, namely the response of the edge to stronger and shorter wavelength excitations.

    \subsection{Dynamical structure factor}
    \label{subsection:IV.A}
\begin{figure}[h]
        \begin{adjustbox}{width=0.48\textwidth, totalheight=\textheight-2\baselineskip,keepaspectratio,right}
        	\includegraphics[]{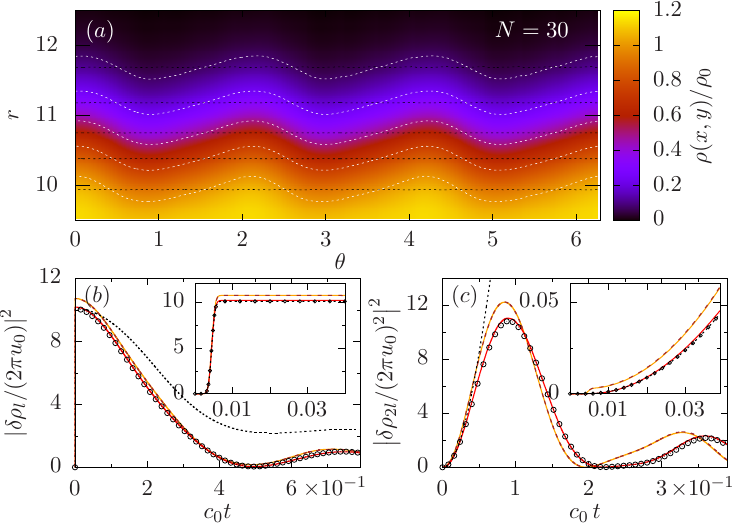}
        \end{adjustbox}
        \caption{(a) Colorplot of the density near the edge at $c_0 t\simeq0.1$ after 
        an excitation pulse of the form \eqref{eq:Ul} with an intensity large enough to induce a significant non-linear dynamics on this temporal scale. White (black) lines are iso-density contours for the 
        excited (unexcited) system.
        (b,c) Time-evolution of the fundamental and second harmonic spatial Fourier components of the edge density variation of $N=30$ (red) and $N=9$ (yellow) clouds. ED data for $N=9$ are shown as brown dashed lines as a benchmark. Dotted black lines and black dots indicate respectively the solution of the semi-classical equation \eqref{eq_NonLinearClassicalHydrodinamicalEquation} and of the quantum model $\hat{H}_{\chi LL}^{NL}$. Insets show a magnified view of the dynamics at early times. Same trap parameters as in Fig.\ref{fig_GSDensity_Spectrum}, filling factor $\nu=1/2$.
        \label{fig_NonLinearDynamics}
    	}
    \end{figure}

    As we have seen in the previous Section, anharmonic confinements cause the DSF to broaden [Fig.\ref{fig_bulkCurrent_SF_DSF_FreqShift}(b)] within a finite frequency window, whose extension turns out (\ref{appendix:D}) to be proportional to $l^2$ and to the 
    curvature of the trap potential at the classical radius
    \begin{equation}
        \label{eq:curvature_angular}
        c_0=\left.R_{cl}^{-1}\partial_r\left({r}^{-1}\partial_r V_\text{conf}(r)\right)\right|_{R_{cl}}=\lambda\,\delta(\delta-2)R_{cl}^{\delta-4}\,,
    \end{equation}
    a quantity related to the second $l$-derivative of the LLL projection of $V_{\rm conf}(r)$, which physically corresponds to the radial gradient of the angular velocity.
    Like in the IQH case~\cite{ourepl}, the broadening is responsible for the 
    decay of the oscillations at late time that is visible in Fig.\ref{fig_NonLinearDynamics}(b).
    
    However, in contrast to the IQH case, the distribution of the DSF
    weights 
    at fixed angular momentum $l$ is non-flat: as one can see in Fig. \ref{fig_bulkCurrent_SF_DSF_FreqShift}(b), within each $l$ sector, the weight of the states close to the high-energy threshold is suppressed, while the one of the states close to the low-energy threshold is reinforced. This behavior is in close analogy to what was found for a fermionic LL beyond the linear dispersion approximation~\cite{Imambekov,Glazman,Pustilnik,Price} and will be the subject of further investigation~\cite{NardinPRA}.

    \begin{figure}[h]
    	\begin{adjustbox}{width=0.48\textwidth, totalheight=\textheight-2\baselineskip,keepaspectratio,right}
        	\includegraphics[]{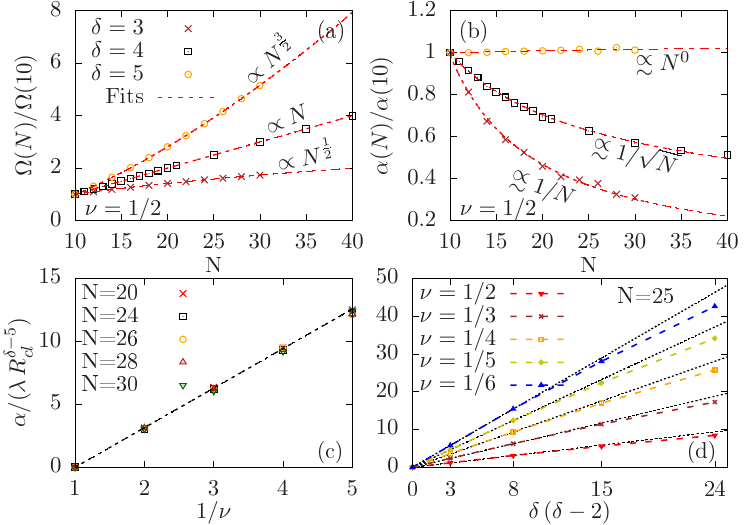}
        \end{adjustbox}
        \caption{(a,b) Normalized angular velocity $\Omega$ and group velocity dispersion parameter $\alpha$ as a function of $N$ for different trap exponents $\delta$ at a constant $\nu=1/2$.
        (c,d) Normalized $\alpha$ as a function of inverse filling $1/\nu$ for (c) $\delta=4$ and different $N$, and (d) as a function of trap curvature $\propto \delta(\delta-2)$ for different fillings $\nu$ at given $N$. All points are extracted from low-$l$ fits to the numerical MC predictions for $\omega_l$ as a function of $l$.         \label{fig_scalings}}
    \end{figure}   
    
   \subsection{Group velocity dispersion}
   \label{subsection:IV.B}
    This asymmetrical distribution of the DSF makes its center-of-mass frequency shift from the low-energy result $\omega_l\simeq \Omega\,l$.
    EE experience a wavevector-dependent frequency-shift and, thus, a finite group velocity dispersion.
    As shown in Fig. \ref{fig_bulkCurrent_SF_DSF_FreqShift}(d), the negative shift gets stronger according to a cubic law at small $l$, 
    \begin{equation}
        \label{eq:modified_dispersion_linear_waves}
        \omega_l=\Omega\,l -\alpha\, l^3\,.
    \end{equation} 
    Note that the cubic form of the frequency shift is different from the quadratic term that appears in typical non-chiral Luttinger liquid theories describing, e.g., interacting Fermi gases, as well as from  the Benjamin-Ono term introduced in the context of FQH fluids in~\cite{Wiegman, Wiegman2} and already critically scrutinized on the basis of conformal field theory and symmetries in~\cite{FernSimon}.
    
    Whereas the results in Fig.\ref{fig_bulkCurrent_SF_DSF_FreqShift}(d) may suggest that the shift is a finite-size effect, a careful account of the $N$ dependence, of the geometry and confinement parameters indicates that the effect persists in the macroscopic limit.
    To this purpose, we note that as $N$ increases at fixed trapping parameters $\lambda,\delta$, the cloud gets correspondingly larger as $R_{cl}=\sqrt{2N/\nu}$, so the effective spatial wavevector of an excitation at $l$ decreases as $q=l/R_{cl}$. 
    At fixed $q$, we expect the frequency shift to be proportional to the curvature of the confining potential in a straight-edge geometry, which 
    in our case suggests
    \begin{equation}
        \label{eq:linear_waves_dispersion}
       \alpha \, l^3 = \beta_\nu \,\tilde{c}_0\, q^3,
    \end{equation}
    with
    \begin{equation}
        \label{eq:c_tilded}
        \tilde{c}_0=R_{cl}^2\,c_0=\lambda\,\delta(\delta-2)\,R_{cl}^{\delta-2} 
    \end{equation}
    and a size-independent $\beta_\nu$.
    This functional form is validated against the numerical results in Fig.\ref{fig_scalings}(b-d). 
    Panel (b) shows that $\alpha$ is indeed proportional to $\sqrt{N}^{\,\delta-5}$ at fixed $\lambda$.
    Panels (c,d) illustrate the linear dependence on the filling factor and on the trap curvature parameter, respectively. 
    
    From these data, we extract a macroscopic coefficient~\footnote{Note that additional, yet typically smaller and opposite in sign group velocity dispersion effects may arise from higher-order terms in the single-fermion dispersion even for $\nu=1$~\cite{ourepl}.}
    \begin{equation}
        \label{eq:beta_nu}
        \beta_\nu \simeq \frac{\pi}{8}\frac{1-\nu}{\nu}.
    \end{equation}
    Since $\beta_\nu\propto1-\nu$, we see that the frequency shift \eqref{eq:linear_waves_dispersion} is related to the strong correlations of the quantum liquid, for it vanishes at integer filling.
    Work is in progress to understand this result in connection with the Hall viscosity\cite{Avron} and the 
    magneto-roton excitations in the bulk
    of the FQH fluid\cite{GMP}.
    
\subsection{Non-linear dynamics}
\label{subsection:IV.C}
    When the excitation strength increases, nonlinear effects start to play an important role in the edge mode evolution. Numerical results illustrating this physics are displayed in Fig.\ref{fig_NonLinearDynamics}: panel (a) shows the
    density profile of the cloud edge after a relatively long evolution time past a sinusoidal 
    excitation with given $l$. 
    In contrast to the weak excitation case discussed above where the density profile keeps at all times a plane-wave form proportional to $\cos(l\theta - \omega_l t)$, here a marked forward-bending of the waveform is visible, leading to a sawtooth-like profile. Upon angular Fourier transform, this asymmetry corresponds to the appearance of higher spatial harmonics.
    
    The physical mechanism underlying the nonlinearity can be understood in analogy with the IQH case~\cite{ourepl}. Because of the incompressibility condition, a local variation $\delta\rho(\theta)$ of the radially-integrated angular density  must correspond to a variation of the cloud radius $\delta R(\theta) \simeq \delta\rho(\theta)/ (\rho_0 R_{cl})$. This then leads to a variation of the local angular velocity \begin{equation}
    \bar{\Omega}(\theta)=\left.{r}^{-1}{\partial_r V_\text{conf}}\right|_{R(\theta)} \simeq \Omega + ({2\pi c_0}/{\nu})\,\delta\rho\,.    
    \end{equation}
    This nonlinear effect can be combined with the group dispersion and the 
    excitation potential $U(\theta,t)$ discussed above into a single semiclassical evolution equation. 
  
  For simplicity, we formulate the equation in terms of the 1D density variation
  along a ``straightened" edge. By back-substitution, obtaining the one in terms of angular parameters is straightforward.
    $\sigma(\zeta,t)=\delta\rho(\theta,t)/R_{cl}$, with $\zeta = R_{cl}\theta$ being the physical position along the edge. The resulting evolution equation
    \begin{equation}
        \label{eq_NonLinearClassicalHydrodinamicalEquation}
        \frac{\partial \sigma}{\partial t} =
        -\left[v_0 + \frac{2\pi \tilde{c}_0}{\nu}\sigma\right] \frac{\partial\sigma}{\partial\zeta}
        -\beta_\nu \tilde{c}_0 \frac{\partial^3\sigma}{\partial\zeta^3}
        -\frac{\nu}{2\pi}\frac{\partial U}{\partial \zeta}
    \end{equation}
    has the form of a driven classical KdV equation~\cite{KdV,KdV2} whose coefficients only involve macroscopic parameters such as the linear speed
    \begin{equation}
        \label{eq:velocity_linear_edge}
        v_0=R_{cl}\,\Omega
    \end{equation}
    determined via \eqref{eq:velocity} by the transverse response to the inward trapping force at the cloud edge, $v_0\sim -\left.\partial_r V_{\text{conf}}(r)\right|_{R_{cl}}$. The confinement potential curvature is defined via Eqs.\eqref{eq:curvature_angular} and \eqref{eq:c_tilded} and is proportional to the second derivative $\tilde{c}_0\sim \left.\partial^2_r V_{\text{conf}}(r)\right|_{R_{cl}}$, namely the gradient of the trapping force.
    
    As one can see in the time evolution of the spatial Fourier components of the density shown in Fig.\ref{fig_NonLinearDynamics}(b,c), the semiclassical equation accurately reproduces the numerical evolution up to relatively long times, where the forward-bending due to the density dependent speed of sound is well visible.
    At later times, the broadening of the DSF discussed above starts to play a dominant role, giving rise to the collapse and revival features visible in the plots. 
    
\section{Non-linear chiral Luttinger liquid theory}
\label{section:V}
    In order to properly capture these last features, quantum effects must be included in the theoretical description. In this perspective, the semiclassical evolution \eqref{eq_NonLinearClassicalHydrodinamicalEquation} can be seen as the classical limit of the Heisenberg equation for the density operator of a $\chi$LL supplemented with a group velocity dispersion term and a forward-scattering non-linearity. 
    
    This reasoning suggests the following form for the lowest non-universal corrections to the quantum $\chi$LL Hamiltonian for our FQH fluid,
    \begin{multline}
        \label{eq_NonLinearXLL}
        \hat{H}_{\chi\text{LL}}^{NL}= \int\!d\zeta\,
        \left[ \frac{\pi\,v_0}{\nu}\,\hat{\sigma}^2-
        \frac{\pi\,\beta_\nu \tilde{c}_0}{\nu}\!\left(\frac{\partial\hat{\sigma}}{\partial\zeta}\right)^2 +\right. \\
    +   \left. \frac{2\pi^2 \tilde{c}_0}{3\nu^2}\hat{\sigma}^3+ U(\zeta,t)\,\hat{\sigma}\right]
    \end{multline}
    where the the density operator of the chiral edge mode obeys the usual $\chi$LL commutation rules~\cite{Wen_topo,Tong},
    \begin{equation}
        \left[\hat{\sigma}(\zeta),\hat{\sigma}({\zeta'})\right]=-i\,\frac{\nu}{2\pi}\,\partial_\zeta\delta(\zeta-\zeta')\,.
        \label{eq:commutator}
    \end{equation}
    
    It is straightforward to verify that the evolution equation for $\hat{\sigma}$ that is obtained by taking the classical limit of the Heisenberg equation
    \begin{multline}
        \label{eq_NonLinearQuantumHydrodinamicalEquation}
        \frac{\partial \hat{\sigma}}{\partial t} =i\left[\hat{H}_{\chi\text{LL}}^{NL},\hat{\sigma}\right]=\\
        =- v_0\frac{\partial\hat{\sigma}}{\partial\zeta} 
        - \frac{\pi \tilde{c}_0}{\nu} \frac{\partial(\hat{\sigma}^2)}{\partial\zeta}
        - \beta_\nu \tilde{c}_0 \frac{\partial^3\hat{\sigma}}{\partial\zeta^3}
        -\frac{\nu}{2\pi}\frac{\partial U}{\partial \zeta}
    \end{multline}
    indeed recovers the classical wave equation \eqref{eq_NonLinearClassicalHydrodinamicalEquation} when operators are replaced by complex numbers.
    
    The different terms in the Hamiltonian \eqref{eq_NonLinearXLL} correspond to the different physical effects discussed in the previous Sections. The first term, proportional to $\hat{\sigma}^2$, is quadratic in the density operators $\hat{\sigma}$: it is already present in the standard chiral Luttinger liquid Hamiltonian and accounts for the increase of energy of the cloud when the edge is deformed from its equilibrium position.
        The second term is proportional to the second spatial derivative $(\partial_\zeta \hat{\sigma})^2$ and is still quadratic in the density operators: it arises from the cubic correction to the dispersion of weak-amplitude waves in \eqref{eq:modified_dispersion_linear_waves}; the additional third derivative appearing in the corresponding term in the wave equation \eqref{eq_NonLinearClassicalHydrodinamicalEquation} comes from the derivative present in the commutator \eqref{eq:commutator}. The microscopic origin of this term will be the subject of future work -- here we just note that it has the suggestive form of a surface-tension energy.
        
        The third term is proportional to $\hat{\sigma}^3$ and therefore is no longer quadratic in $\hat{\sigma}$: it stems from the intrinsic nonlinearities discussed in \ref{subsection:IV.C} and it describes interactions among the bosonic modes of the chiral Luttinger liquid.
Finally, the last term proportional to the density operator $\hat{\sigma}$ is analogous to the coupling to the electromagnetic field in the standard $\chi$LL theory~\cite{Wen_topo, Wen2}: in our model, it describes the external driving generated by the coupling of the cloud density to the external potential $U(\theta, t)$ in Eq.\eqref{eq:Ul}.

    All numerical coefficients appearing in the quantum Hamiltonian \eqref{eq_NonLinearXLL} can be straightforwardly calculated in terms of the FQH filling $\nu$ and the radial dependence of the confinement potential $V_{\rm conf}(r)$ around the classical radius $r=R_{cl}=\sqrt{2N/\nu}$ using Eqs.
    \eqref{eq:c_tilded}, \eqref{eq:beta_nu}, and \eqref{eq:velocity_linear_edge}. 
    This confirms the physical expectation that the edge dynamics only depends on the local features of the confinement. The resulting formulas
    \begin{eqnarray}
        v_0&=&  \left.\partial_r V_{\rm conf}(r)\right|_{R_{cl}}\\
        \tilde{c}_0&=& \left.R_{cl}\partial_r\left(r^{-1}\partial_r V_{\rm conf}(r)\right)\right|_{R_{cl}} 
    \end{eqnarray}
    can be used to obtain quantitative predictions for specific physical systems.

    The surprisingly good accuracy of the physical predictions of the nonlinear  $\chi$LL Hamiltonian \eqref{eq_NonLinearXLL} is showcased in 
Fig.\ref{fig_GSDensity_Spectrum}(d),
Fig.\ref{fig_bulkCurrent_SF_DSF_FreqShift}(b) and Fig.\ref{fig_NonLinearDynamics}(b,c) for the eigenenergy spectrum, the DSF~\footnote{The slight deviations at large momenta are due to the higher-order group velocity dispersion and finite-size effects that are visible also in Fig.~\ref{fig_bulkCurrent_SF_DSF_FreqShift}(c,d).} and the complete time evolution, respectively. In each of these plots, the predictions of \eqref{eq_NonLinearXLL} are compared to the result of the full microscopic Hamiltonian and an excellent agreement is found. An analogous agreement is shown in Appendix \ref{appendix:E} for additional observables.

All together, these results strongly support the predictive power of the nonlinear $\chi$LL model. Given the favourable scaling of its numerical complexity with particle number $N$ as compared to the full two-dimensional calculations, the one-dimensional nonlinear $\chi$LL appears as a most promising tool to describe the dynamics of large FQH clouds well beyond the limitations of the full many-body description. 

    
    \section{Experimental observability}
    \label{section:VI}
    We conclude the work with a brief discussion of the actual relevance of our predictions in view of experiments with synthetic quantum matter systems, in particular trapped atomic gases for which an artillery of experimental tools is already available. 
    
    As several strategies to induce synthetic magnetic fields are nowadays well established, from rotating traps~\cite{Bretin, Schweikhard} to combinations of optical and magnetic fields~\cite{TopoBandsForAtoms, review_Goldman_synthetic}, the open challenge is to reach sufficiently low atomic filling factors and sufficiently low temperatures to penetrate the fractional quantum Hall regime~\cite{ReviewNigel,review_dalibard}: an intense work is being devoted to this issue from both the theoretical~\cite{Andrade} 
    and experimental sides, and promising preliminary observations have appeared in the literature~\cite{Gemelke,Zwierlein2,Leonard}. In this, an important challenge is to design an adiabatic protocol for reaching a Laughlin state with large fidelity.
    Once the desired many-body state is generated, arbitrary confinement potentials can be generated with optical techniques~\cite{Bretin,Gaunt} and the response to rotating potentials of the form \eqref{eq:Ul} can be measured via the same tools used, e.g., to study surface excitations of rotating superfluid clouds~\cite{dalibard}.
    
    Most remarkably, we have shown in Fig.\ref{fig_bulkCurrent_SF_DSF_FreqShift} that this measurement provides a precise measurement of the transverse conductivity already for moderate cloud sizes $N\sim 10$. This suggests that a smoking gun of the topological nature of the many-body state can be obtained in strongly correlated atomic clouds with realistic sizes trapped in fast rotating potentials~\cite{review_dalibard}.
    
    While transverse conductivity features are independent of the shape of the confinement potential, both the group velocity dispersion and the nonlinear effects crucially depend on the trap anharmonicity that also helps stabilizing the cloud at large rotation speeds close to the centrifugal limit. A rough estimate of the maximum potential curvature $\tilde{c}_0$ that the FQH liquid can stand before being significantly affected is set by the many-body gap over the squared magnetic length. Since both the group velocity dispersion and the nonlinearity terms in (\ref{eq_NonLinearClassicalHydrodinamicalEquation}-\ref{eq_NonLinearXLL}) scale proportionally to the curvature $\tilde{c}_0$ and the chiral dynamics factors out as a rigid translation at $v_0$, such an upper bound on $\tilde{c}_0$ does not impose any restriction on the observability of interesting effects due to their interplay. It only requires that the dynamics is followed on a temporal scale much longer than the inverse many-body gap, a condition which is anyway automatically enforced upon working with a correlated many-body state.
    
    To be more specific, let us consider again the case of ultracold bosons in the fast rotation regime already mentioned in Sec.\ref{subsection:II.A}. In this case, for a $\delta=4$ quartic anharmonic potential,
    the curvature parameter can be written as $c/2\Omega_r = \lambda \hbar/M^2\Omega_r^3$.
    Based on the constraints discussed above, the timescale for the correction of linear waves \eqref{eq:modified_dispersion_linear_waves} is then set by the reciprocal of $T_l =1/(c l^3)$, which is $10^2\div10^3$ longer than the timescale set by the many-body energy gap $\Delta$. In order to be able to observe the correlated many-body state one needs to maintain the system over a timescale much longer than the reciprocal of the many-body energy gap.
    
    When the droplet gets excited by a time-dependent external potential of the form \eqref{eq:Ul}, the edge density variation $\delta\sigma$ predicted by Eq.\eqref{eq_NonLinearQuantumHydrodinamicalEquation} 
    in the linear regime of weak excitations is proportional to $l u_0 \tau/R_\text{cl}$, where $\tau$ is the duration of the (short) Gaussian excitation pulse and $u_0$ is its strength~\footnote{Of course, stronger excitations could be obtained by a careful engineering of the excitation sequence that may include resonant temporal oscillations.}. 
    Given the incompressible nature of the FQH fluid, the edge density variation $\delta\sigma$ then results in a corresponding variation of the cloud radius $\delta R / R_\text{cl} = \delta\sigma\, \nu/(2\pi \, R_\text{cl}) \sim l u_0 \tau \nu / N$ which can be detected either in-situ or, if needed, after a time-of-flight expansion: as one can see in Fig.\ref{fig_NonLinearDynamics}, the relative change in the cloud radius can be a significant fraction of its equilibrium value, which supports the experimental observability of our predictions.
    
    On-going work is addressing the robustness of our predictions to different geometries and configurations. On one hand, a forthcoming manuscript~\cite{NardinPRA} will discuss how a similar physics is obtained for macroscopic clouds trapped by in non-smooth hard-wall potentials. On the other hand, ab initio exact diagonalization calculations are presently investigating the edge response of few-body FQH clouds in those lattice geometries that are presently under active experimental investigation~\cite{Leonard}. 
    
    
    A different strategy to observe the nonlinear dynamics and highlight KdV behaviours is to induce a spatially localised density modulation by selectively removing a controlled number of particles in the vicinity of the system edge~\cite{Cooper2}: the study of the dynamics of such  particle-number-non-conserving, spatially localized and strongly nonlinear excitations and of their interplay with the particle-number-conserving edge modes studied in this work will be the subject of future work.
    
    \section{Conclusions}
    \label{section:VII}
    In this work we have reported a numerical study of the linear and nonlinear edge dynamics of a 
    fractional quantum Hall cloud of macroscopic size. Our calculations are based on a novel numerical method based on expanding the many-body wavefunction in the basis of Laughlin states and evaluating the matrix elements of the Hamiltonian and of the main observable quantities in this basis by Monte Carlo techniques. This allows us to follow in time the evolution of the cloud in response to different excitation sequences.
    
    Our calculations highlight a number of effects of direct experimental interest both at linear and nonlinear regime, such as a sizable group velocity dispersion of the edge mode and a significant amplitude-dependent response. Since our conclusions are based on a very generic model, they directly apply to fractional quantum Hall fluids both in atomic or photonic synthetic matter and in electronic systems. As such, they are prone to experimental investigations with state-of-the-art systems.
    
    From the theoretical side, the numerical results are used  to build an effective one-dimensional nonlinear chiral Luttinger liquid ($\chi$LL) quantum formalism describing a dynamics in the form of a quantum Korteweg-de Vries equation. The predictive power of the nonlinear $\chi$LL formalism is successfully validated at a quantitative level by comparing its predictions against the full numerics. As compared to the full two-dimensional calculations, the $\chi$LL approach has a much more favourable scaling with system size, which allows to address macroscopically large systems. 
    
    Work is presently in progress to combine the quantum $\chi$LL formalism with refermionization techniques to understand the peculiar exponents numerically observed in the dynamic structure factor of edge excitations. In the future, this formalism will be a natural starting point to investigate more subtle nonlinear effects in the edge dynamics such as solitonic excitations. Once supplemented with terms describing tunneling processes between FQH edges~\cite{ZenoMSc}, it holds great promise in view of using FQH fluids as a novel platform for nonlinear quantum optics of edge excitations and highlight observable signatures of the anyonic statistics of FQH excitations.

\begin{acknowledgements}
We acknowledge financial support from the H2020-FETFLAG-2018-2020 project ``PhoQuS'' (n.820392). IC acknowledges financial support from the Provincia Autonoma di
Trento and from the Q@TN initiative. Continuous discussions with Elia Macaluso, Zeno Bacciconi and Daniele De Bernardis are warmly acknowledged.
\end{acknowledgements}

\appendix


\section{Statistics of the sampling}
\label{appendix:A}
    In order to estimate the statistical error of the Monte Carlo sampling, we performed some statistical analysis on the numerical data. In particular, we split the calculations of our observables into $M=250$ groups for the same droplet configuration. The obtained results are treated as a population of which we studied the statistics.
    \begin{figure}[htbp]
    	\begin{adjustbox}{width=0.48\textwidth, totalheight=\textheight-2\baselineskip,keepaspectratio,right}
        	\includegraphics[]{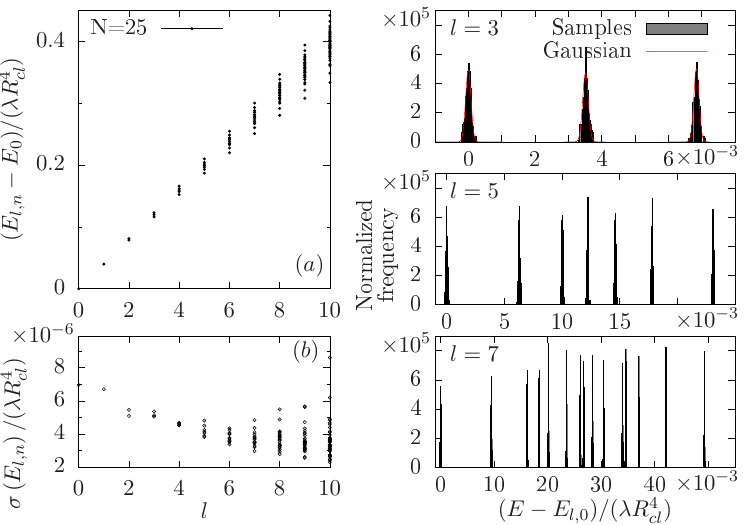}
        \end{adjustbox}
        \vspace{0.0cm}\caption{
        (a) Eigenenergy spectrum (with errorbars) for a $N=25$, $\nu=1/2$ FQH cloud confined by a $\delta=4$ quartic potential.
        (b) Magnified view on the statistical errors on the eigenenergies.
        The panels on the right show histograms for the $M=250$ Monte Carlo realizations of the energy spectrum in each  $l$-sector. Each point is obtained by an independent run.
        \label{fig_EnergiesStatistics}}
    \end{figure}
    \begin{figure}[htbp]
    	\begin{adjustbox}{width=0.48\textwidth, totalheight=\textheight-2\baselineskip,keepaspectratio,right}
        	\includegraphics[]{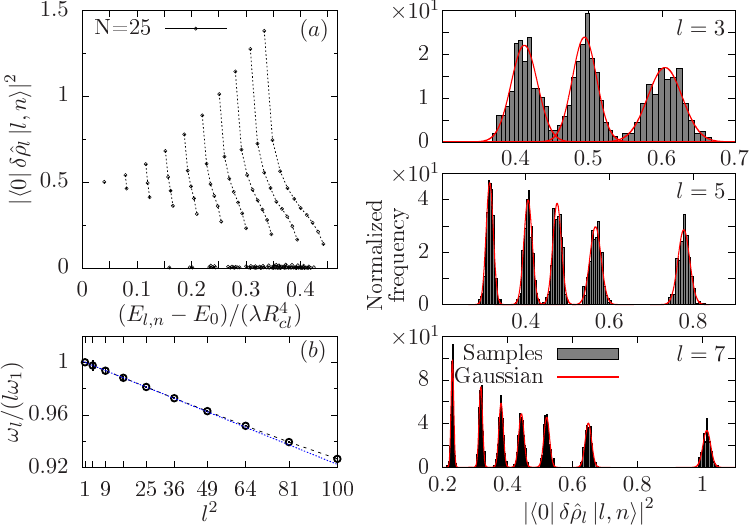}
        \end{adjustbox}
        \vspace{0.0cm}\caption{        
        (a) DSF weights $|\bra{0}\delta\hat{\rho}_l\ket{l,n}|^2$ (with errorbars) for a $N=25$, $\nu=1/2$ FQH cloud confined by a $\delta=4$ quartic potential.
        (b) Suitably normalized first moment $\omega_l$ of the DSF (with errorbars).
    The panels on the right show histograms for the $M=250$ Monte Carlo realizations of the DSF weights in each $l$-sector.
        \label{fig_DSFStatistics}}
    \end{figure}        
     
    The average energies 
    \begin{equation}
        E_{l,n}=\frac{1}{M}\sum_{i=1}^M e_{l,n}[i]
    \end{equation}
    are shown in Fig. \ref{fig_EnergiesStatistics}(a) with their standard errors 
    \begin{equation}
        \sigma(E_{l,n})=\left(\frac{1}{M(M-1)}\sum_{i=1}^M\left(e_{l,n}[i]-E_{l,n}\right)^2\right)^{\frac{1}{2}}.
    \end{equation}
    Since these latter are very small and almost invisible on panel (a), we have replotted them separetely in panel (b). Histograms of the $M=250$ samples for the eigenstate energies at a few values of $l$ are shown in the right panels.
    
    The same analysis has been repeated for the DSF; the results for the DSF weights are shown in Fig. \ref{fig_DSFStatistics}(a).  Again, the error bars are too small to be seen by eye on that scale. Histograms of the $M=250$ samples for a few $l$ components of the DSF are shown in the right panels.
    Error propagation then yields small but sizeable errorbars on the central frequency $\omega_l$, in particular at $l=1$, as shown in panel (b).

\section{Excitations with a radial dependence}
\label{appendix:B}
    \begin{figure}[htbp]
    	\begin{adjustbox}{width=0.48\textwidth, totalheight=\textheight-2\baselineskip,keepaspectratio,right}
        	\includegraphics[]{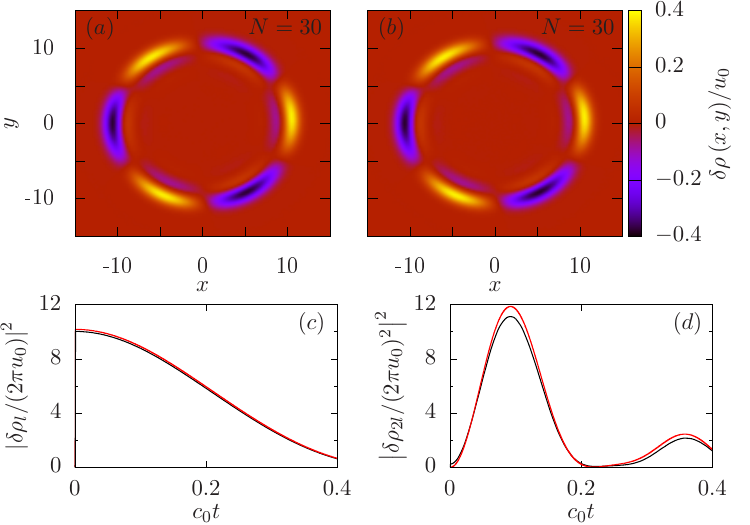}
        \end{adjustbox}
        \vspace{0.2cm}\caption{
        Density variation $\delta\rho(x,y)$ at a time $c_0 t\simeq 0.1$ after the $\nu=1/2$ cloud has been excited by means of (a) 
        an excitation with a non-trivial radial profile described by \eqref{eq:fullcoupling} or (b) by a radially flat profile.
        (c,d) Time-evolution of the fundamental and second harmonic spatial Fourier components of the edge density variations in the same two cases (black and red).
        \label{fig_radialExcitation}}
    \end{figure}
    
The picture presented in the main text remains valid under reasonable approximations even when the externally applied excitation depends on the radial coordinate.
The external potential couples to the density (apart for a time-dependent additive constant which is anyway irrelevant for the dynamics) via
    \begin{equation}
        \label{eq_fullCoupling}
        \hat{V}(t) = \int U(\mathbf{r};t)\,\delta\hat{\rho}(\mathbf{r})\,d^2\mathbf{r}.
    \end{equation}
    For edge excitations, the support of the density variation $\delta\hat{\rho}(\mathbf{r})$ is exponentially localized near the edge, $r\simeq R_\text{cl}$: if the excitation is constant over the width of the edge mode, we can approximate
    \begin{multline}
        \label{eq_azimuthalCoupling}
        \hat{V}(t) \simeq \int U(R_\text{cl},\theta;t)\left(\int \delta\hat{\rho}(\mathbf{r})\,r\,dr\right)\,d\theta = \\ 
       = \int  U(R_\text{cl},\theta;t)\,\delta\hat{\rho}(\theta)\,d\theta \,,
    \end{multline}
    which indeed yields a minimal coupling between the edge density variation and an effectively azimuthal excitation.
    For this formula to remain valid for a radially-dependent potential, we can expect that the potential $U$ has to reach the bulk on one side and overlap with the whole edge on the other side. 
    {This condition is needed for the quantized transverse Hall current to flow from the bulk towards the edge during the excitation time, so that the edge density variation is proportional to the macroscopic bulk transverse conductivity set by the filling fraction.}

    To validate this physical picture, we compare the calculations presented in the main text for a radially constant potential with analogous calculations with an excitation of the form
    \begin{equation}
        U(r,\theta;t)=U_l(t)\,\left(r/R_\text{cl}\right)^l\,e^{i l\,\theta} + \textrm{c.c.}
        \label{eq:fullcoupling}
    \end{equation}
    for which the radial variation of the excitation potential over the edge-mode shape may be not negligible. As shown in Fig.\ref{fig_radialExcitation}, {good} qualitative agreement with the results for a flat $U(\theta;t)=U(R_\text{cl},\theta;t)$ is found: the density variations $\delta\rho(x,y)$ are in fact practically indistinguishable. Note that the excitation considered here was strong enough to trigger visible non-linear effects.
    
    The comparison has been made more quantitative by looking at the time-evolution of the spatial Fourier transforms of the edge density (bottom panels). The fundamental mode in the two cases can hardly be told apart. Slight quantitative differences appear in the second spatial harmonic, even though the qualitative shape remains the same. This confirms that the approximation made in \eqref{eq_azimuthalCoupling} is a good one, especially at small $l$, so the simpler form \eqref{eq_azimuthalCoupling} is an accurate effective description also for the more general coupling \eqref{eq_fullCoupling}.

\section{Linear response within the $\chi$LL theory}
\label{appendix:C}
The key observable we consider is the edge density variation defined as
\begin{equation}
    \delta\hat{\rho}(\theta) = 
    \int_0^\infty \left(
    \hat{\psi}^\dagger(\mathbf{r})\hat{\psi}^{\vphantom{\dagger}}(\mathbf{r})-
    \braket{\hat{\psi}^\dagger(\mathbf{r})\hat{\psi}^{\vphantom{\dagger}}(\mathbf{r})}
    \right)r\,dr
\end{equation}
where the bra-kets denote the expectation value on the ground state and $\psi^\dagger(\mathbf{r})$ is the particle-creation operator at position $\mathbf{r}$.

Within linear response theory, the edge density variation induced by the external perturbing potential of the form \eqref{eq_azimuthalCoupling} reads
\begin{equation}
    \label{eq_Kubo}
    \braket{\delta\tilde{\rho}(\theta,t)}=-i
    \Braket{
    \left[\delta\tilde{\rho}(\theta,t),\int_{-\infty}^t\tilde{V}(t')dt'\right]
    }
\end{equation}
where the system is assumed to be initially in its ground state at $t\rightarrow-\infty$, higher order terms $\mathcal{O}(U^2)$ have been neglected and the tilde indicate interaction picture with respect to the unperturbed $U=0$ Hamiltonian.

With straightforward algebra, the above formula can be rewritten as
\begin{multline}
    \label{eq_Kubo2}
    \braket{\delta\hat{\rho}(\theta,t)}=2\,\Im\int_{-\infty}^t dt'\int d\theta' \,\,U(\theta',t')\\
    \Braket{
    \delta\hat{\rho}(\theta)\,e^{-i(\hat{H}-E_0)(t-t')}\,\delta\hat{\rho}(\theta')
    }.
\end{multline}
Introducing the Fourier transforms
\begin{equation}
    \begin{cases}
        \delta\hat{\rho}(\theta) = \frac{1}{2\pi}\sum_{l\neq0} e^{i l\theta} \delta\hat{\rho}_l    
        \\
        U(\theta, t) = \frac{1}{2\pi}\sum_{l\neq0} e^{i l\theta} U_l(t)\,    
    \end{cases}
\end{equation}
this can be reformulated as 
\begin{multline}
    \label{eq_Kubo3}
    \braket{\delta\hat{\rho}(\theta,t)}=\\
    =2\,\Im\int_{-\infty}^t dt'\frac{1}{(2\pi)^2} \sum_{l\neq0} e^{i l\theta}\,U_l(t')\,\mathcal{C}_l(t-t')
\end{multline}
where the rotational invariance of the ground state has been used to remove a summation, $\mathcal{C}_{ll'}=\mathcal{C}_{l}\,\delta_{ll'}$ with
\begin{equation}
    \label{eq_correlation}
    \mathcal{C}_l(t)=
    \Braket{
    \delta\hat{\rho}_l \,e^{-i(\hat{H}-E_0)t}\, \delta\hat{\rho}_{-l}
    }\,.
\end{equation}
If we are interested in the late time dynamics of the system once the perturbation pulse has gone ($U_l(t)\to 0$ for late times), we can replace the upper boundary of the time integral with $t\rightarrow\infty$, use the convolution theorem and write
\begin{equation}
    \label{eq_Kubo4}
\braket{\delta\hat{\rho}(\theta,t)}=\frac{1}{\pi}\Im\left[
    \sum_l e^{i l \theta}
    \int \widetilde{U}_{l}(\omega) \,S_l(\omega) \,e^{-i\omega t}\,d\omega \right]
\end{equation}
where
\begin{eqnarray}
    \widetilde{U}_{l}(\omega) = \int \frac{dt}{2\pi}\, e^{i\omega t}\, U_l(t)
        \\
    S_l(\omega)=\int \frac{dt}{2\pi}\,e^{i\omega t} \,\mathcal{C}_l(t). \label{eq:S_l}
\end{eqnarray}
Combining \eqref{eq:S_l} with \eqref{eq_correlation} allows to recover the edge dynamic structure factor. As long as the confinement and 
excitation potentials are weak enough not to excite states above the many-body gap, we can introduce a projector onto these states only and rewrite

\begin{equation}
    \label{eq_laughlin_projected_DSF}
    S_l(\omega)=\sum_n\,\delta\left(\omega-\omega_{l,n}\right) \left|\bra{0}
    \delta\hat{\rho}_l \ket{l,n}\right|^2
\end{equation}
where $\ket{0}$ is the Laughlin ground state and $\omega_{l,n}=E_{l,n}-E_0$ the excitation energy of state $\ket{l,n}$ with respect to the ground state.

Integrating over the frequencies in \eqref{eq_laughlin_projected_DSF} (restriction to energies below the many-body gap is automatically enforced by the projector onto the low-energy subspace) one obtains the edge static structure factor
\begin{equation}
    \label{eq_laughlin_projected_SSF}
    S_l=\sum_n \left|\bra{0}
    \delta\hat{\rho}_l \ket{l,n}\right|^2
\end{equation}
which is invariant under a deformation of the many-body Hamiltonian as long as the gap is not closed, so that a unitary transformation between the ``new" eigenstates $\ket{l,n}'$ and the ``old" ones $\ket{l,n}$ is well defined. Hence, in the long wavelength/low energy limit the edge static structure factor maintains its $\chi$LL value, namely $S_l=\nu l$ when $l\geq0$ and $0$ otherwise, reflecting the chirality of the system. 

Assuming a narrowly peaked DSF at $\omega\simeq\omega_l$ and including the $\chi$LL form of $S_l$, we can approximate \eqref{eq_Kubo4} as
\begin{equation}
    \label{eq_Kubo5}
    \braket{\delta\hat{\rho}(\theta,t)}=-\frac{\nu}{\pi}\frac{\partial}{\partial \theta}
    \sum_{l>0} \Re\left[ e^{i l (\theta-\omega_l t)}
    \widetilde{U}_{l}(\omega_l) \right]\,:
\end{equation}
this formula explicitly displays the proportionality of the edge response to the FQH filling factor and is the key of our proposed measurement scheme of the transverse conductivity.
Of course, this formula
is only valid up to not-too-large times, namely as long as the DSF broadening is not resolved, $\Delta E_l t \ll 1$. 

Note finally that the solution of the semiclassical equation introduced in the main text [Eq.(1) there] perfectly matches this result as long as the nonlinear velocity term can be neglected.

\section{Broadening of the dynamical structure factor of edge modes}
\label{appendix:D}
    \begin{figure}[htbp]
    	\begin{adjustbox}{width=0.48\textwidth, totalheight=\textheight-2\baselineskip,keepaspectratio,right}
        	\includegraphics[]{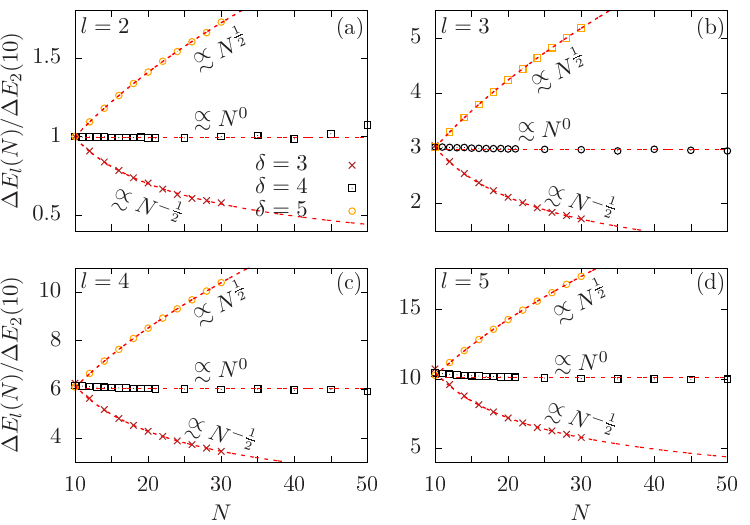}
        \end{adjustbox}
        \vspace{0.0cm}\caption{
        Suitably normalized width $\Delta E_l$ of the DSF of a $\nu=1/2$ FQH cloud as a function of the particle number $N$ for different values of $l=2$ (a), $3$ (b), $4$ (c), $5$ (d). The dashed lines are power law fits to the data and highlight the scaling with $N$ at fixed $l$, for different confinements $\delta$. The fitted exponents are in close agreement with the expected ones indicated in the legends.
        \label{fig_DSFBroadening}}
    \end{figure}
     \begin{figure}[htbp]
    	\begin{adjustbox}{width=0.48\textwidth, totalheight=\textheight-2\baselineskip,keepaspectratio,right}
        	\includegraphics[]{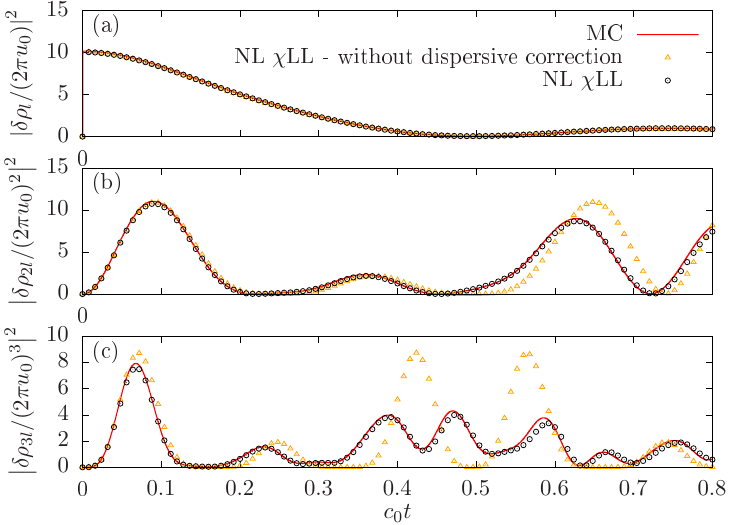}
        \end{adjustbox}
        \vspace{0.0cm}\caption{
        Time evolution of the fundamental (a), second (b) and third (c) harmonic of the spatial Fourier transform of the edge density variation of a $N=30$ cloud at filling factor $\nu=1/2$ in a $\delta=4$ quartic trap. The results of the microscopic MC calculations (red lines) are compared with those of the non-linear $\chi$LL model (black circles). For comparison, the result of a non-linear $\chi$LL model without the dispersive contribution is shown as yellow triangles.
        \label{fig_ThirdFourierComponent}}
    \end{figure}         
    
    When the cloud is non-harmonically confined with $\delta\neq 2$, we have seen in the main text that the DSF broadens  within a finite frequency window, whose width can be easily estimated by looking at the difference $\Delta E_l$ between the largest and smallest energies in a given angular momentum $l$ sector. The corresponding states have in fact a non-vanishing DSF weight $|\bra{0}\delta\hat{\rho}_l\ket{l,n}|^2$ and these energies thus correspond to the thresholds of the DSF.
    
    In close analogy to to the IQH, we expect the DSF to broaden $\propto c\, l^2$. Here we verify this scaling. In particular, data in Fig.\ref{fig_DSFBroadening} suggest the following simple form
    \begin{equation}
        \label{eq_DSFbroadening}
        \Delta E_l = \mu_m\,\frac{c}{2}\,l(l-1).
    \end{equation}
    The proportionality $c\propto R_{cl}^{\delta-4}$ is visible from the $N$ dependence in each $l$ sector. Since all data have been normalized by $\Delta E_{l=2}$ (at a fixed number of particles, $N=10$), the proportionality to $l(l-1)/2$ can be instead read out by looking at the first point on $y$-axis. 
    Notice that, apart for the $m$-dependent proportionality factor, the result in \eqref{eq_DSFbroadening} is exactly the same as in the IQH case, where the lower (upper) threshold corresponds to a particle (hole) created just above (below) the Fermi surface.
    
\section{Quantitative comparison between the microscopic dynamics and the non-linear $\chi$LL model Hamiltonian}
\label{appendix:E}
    
    To further support the nonlinear $\chi$LL model, the numerically calculated microscopic time-evolution was compared with the results of the nonlinear $\chi$LL model for different observables. To this purpose, the 
   free parameters of the model have been determined according to the scaling formulas discussed in the text, without any additional fine-tuning.
    In particular, for a $\delta=4$ quartic confinement the angular velocity of the edge modes is set by $\Omega=4\lambda R_\text{cl}^2$; we have a size-independent curvature $c=8\lambda$ which determines both the cubic phonon dispersion shift coefficient $\alpha=\pi\lambda/R_\text{cl}$ and the strength of the nonlinearity.
    
    The time-evolution of the spatial Fourier transform of the edge density variation calculated by the full numerics and by the $\chi$LL model are compared in Fig. \ref{fig_ThirdFourierComponent}. A very good agreement can be seen, which gets slightly worse at larger angular momenta $l$: this small deviation may be caused by a higher-order correction of the phonon dispersion (beyond the cubic term considered here) and by the increasing difficulty in accurately sampling the matrix elements of the 
    excitation Hamiltonian between higher-$l$ subspaces.
    Note that the cubic correction to the phonon dispersion is essential to correctly capture the late-time dynamics, in particular of the harmonic components at $2l$ and $3l$ [see yellow triangles in Fig. \ref{fig_ThirdFourierComponent}]. Of course, the nonlinear terms are even more essential, as they are responsible for the very appearance of a finite amplitude in the harmonic components.


\begin{thebibliography}{}
    \bibitem{Jain} J. K. Jain, ``Composite Fermions", Cambridge University Press. doi:10.1017/CBO9780511607561
    
    \bibitem{Tong} D. Tong, ``Lectures on the Quantum Hall Effect" (2016), available as arXiv: 1606.06687.
    
    \bibitem{HannahTomoki} 
    T. Ozawa, H. M. Price, {\em Topological quantum matter in synthetic dimensions}, Nature Reviews Physics {\bf 1}, 349 (2019).
    
    \bibitem{ReviewNigel} N. Cooper, Adv. Phys. {\bf 57}, 539 (2008).
    
    \bibitem{review_Goldman_synthetic}
    N. Goldman, G. Juzeliunas, P. \"Ohberg and I. B. Spielman, {\em Light-induced gauge fields for ultracold atoms}, Rep. Prog. Phys. {\bf 77}, 126401 (2014).
    
    \bibitem{review_dalibard} I. Bloch, J. Dalibard, W. Zwerger, Rev. Mod. Phys. {\bf 80} 885 (2008).
    
    
    \bibitem{TopoBandsForAtoms} N. R. Cooper, J. Dalibard, I. B. Spielman, Rev. Mod. Phys. {\bf 91} 015005 (2019).
    
    \bibitem{RMPQFL} I. Carusotto and C. Ciuti, {\em Quantum fluids of light}, Rev. Mod. Phys. {\bf 85}, 299 (2013).
    
    \bibitem{Topo} T. Ozawa, H. M. Price, A. Amo, N. Goldman, M. Hafezi, L. Lu, M. C. Rechtsman, D. Schuster, J. Simon, O. Zilberberg, and I. Carusotto, {\em Topological photonics}, Rev. Mod. Phys. {\bf 91}, 015006 (2019).

    
    \bibitem{NPHYS} I. Carusotto, A. A. Houck, A. J. Koll\'ar, P. Roushan, D. I. Schuster, J. Simon, {\em Photonic materials in circuit quantum electrodynamics}, Nature Physics, {\bf 16}, 268-279 (2020).
    
    \bibitem{Paredes} 
    B. Paredes, P. Fedichev, J. I. Cirac, and P. Zoller, {\em 1/2-Anyons in Small Atomic Bose-Einstein Condensates}, Phys. Rev. Lett. {\bf 87}, 010402 (2001).
    
    \bibitem{Cooper2} N. R. Cooper, Steven H. Simon, ``Signatures of Fractional Exclusion Statistics in the Spectroscopy of Quantum Hall Droplets" Phys. Rev. Lett. {\bf 114}, 106802 (2015)
    
    \bibitem{Regnault} 
    N. Regnault and Th. Jolicoeur, {\em Quantum Hall Fractions in Rotating Bose-Einstein Condensates}, Phys. Rev. Lett. {\bf 91}, 030402 (2003).

    \bibitem{Rizzi} M. Roncaglia, M. Rizzi, J. Dalibard, Sci. Rep. {\bf 1}, 43 (2011).
    
    \bibitem{QHWithOptFluxes} N. R. Cooper, J. Dalibard, Phys. Rev. Lett. {\bf 110}, 185301 (2013).
    
    \bibitem{Rougerie} N. Rougerie, S. Serfaty, and J. Yngvason, {\em Quantum Hall states of bosons in rotating anharmonic traps}, Phys. Rev. A 87, 023618 (2013).
    
    \bibitem{Morris} A. G. Morris, D. L. Feder, ``Gaussian Potentials Facilitate Access to Quantum Hall States in Rotating Bose Gases", Phys. Rev. Lett. {\bf 99}, 240401 (2007)

    \bibitem{Goldman1} N. Goldman, J. Dalibard, A. Dauphin, F. Gerbier, M. Lewenstein, P. Zoller, I. B. Spielman, PNAS {\bf 110}, 6736 (2013).

    \bibitem{Goldman} C. Repellin, N. Goldman, {\em Detecting fractional Chern insulators through circular dichroism}, Phys. Rev. Lett. {\bf 122}, 166801 (2019).
    
    \bibitem{Onur} 
    R. O. Umucalilar, E. Macaluso, T. Comparin, and I. Carusotto, {\em Time-of-Flight Measurements as a Possible Method to Observe Anyonic Statistics}, Phys. Rev. Lett. {\bf 120}, 230403 (2018).
    
    \bibitem{Onur2} 
    E. Macaluso, T. Comparin, R. O. Umucalilar, M. Gerster, S. Montangero, M. Rizzi, and I. Carusotto, {\em Charge and statistics of lattice quasiholes from density measurements: A tree tensor network study, }Phys. Rev. Research {\bf 2}, 013145 (2020).
    
    \bibitem{Onur3} R. O. Umucalilar, M. Wouters, I. Carusotto, {\em Probing few-particle Laughlin states of photons via correlation measurements
    }, Phys. Rev. A {\bf 89}, 023803 (2014).
    
    
    \bibitem{Onur4}  
    R.O. Umucalilar, I. Carusotto, {\em Generation and spectroscopic signatures of a fractional quantum Hall liquid of photons in an incoherently pumped optical cavity}, Phys. Rev. A {\bf 96}, 053808 (2017)
    
    
    
    \bibitem{AMDLH} A. Mu\~noz de las Heras, E. Macaluso, I. Carusotto, {\em Anyonic molecules in atomic fractional quantum Hall liquids: a quantitative probe of fractional charge and anyonic statistics}, Physical Review X {\bf 10}, 041058 (2020).
    
        \bibitem{Unal} 
    M. Raciunas, F. N. \"Unal, E. Anisimovas, and A. Eckardt, {\em Creating, probing, and manipulating fractionally charged excitations of fractional Chern insulators in optical lattices}, Phys. Rev. A {\bf 98}, 063621 (2018).
    
    \bibitem{Stern} A. Stern, Ann. Phys. {\bf 323}, 1, 204-249 (2008)
    
    \bibitem{Gemelke} 
    N. Gemelke, E. Sarajlic, S. Chu, {\em Rotating few-body atomic systems in the fractional quantum Hall regime}, arXiv:1007.2677.

    \bibitem{Greiner}  M. E. Tai, A. Lukin, M. Rispoli, R. Schittko, T. Menke, D. Borgnia, P. M. Preiss, F. Grusdt, A. M. Kaufman, and M. Greiner, Microscopy of the interacting Harper-Hofstadter model in the two-body limit, Nature {\bf 546}, 519 (2017).


    \bibitem{Leonard} J. L\'eonard, {\em et al.}, preprint arXiv:2210.10919 (2022) 
    
    
    \bibitem{Roushan} P. Roushan, C. Neill, A. Megrant, Y. Chen, R. Babbush, R. Barends, B. Campbell, Z. Chen, B. Chiaro, A. Dunsworth, A. Fowler, E. Jeffrey, J. Kelly, E. Lucero, J. Mutus, P. J. J. O’Malley, M. Neeley, C. Quintana, D. Sank, A. Vainsencher, J. Wenner, T. White, E. Kapit, H. Neven, and J. Martinis, {\em Chiral ground-state currents of interacting photons in a synthetic magnetic field}, Nat. Phys. {\bf 13}, 146 (2017).

    \bibitem{Clark} L. W. Clark, N. Schine, C. Baum, N. Jia, J. Simon, {\em Observation of
    Laughlin states made of light}, Nature {\bf 582},
    41 (2020).


    
    
    \bibitem{Nayak} C. Nayak, S. H. Simon, A. Stern, M. Freedman, S. Das Sarma, Rev. Mod. Phys. {\bf 80}, 1083 (2008)
    
    \bibitem{DePicciotto} R. de-Picciotto, M. Reznikov, M. Heiblum, V. Umansky, G. Bunin, D. Mahalu, Nature {\bf 389}, 162-164 (1997)
    
    \bibitem{Nakamura} J. Nakamura, S. Liang, G. C. Gardner, M. J. Manfra, Nat. Phys. {\bf 16}, 931-936 (2020)
    
    \bibitem{DeBartolomei} H. Bartolomei, M. Kumar, R. Bisognin, A. Marguerite, J. M. Berroir, E. Bocquillon, B. Plaçais, A. Cavanna, Q. Dong, U. Gennser, Y. Jin, G. Fève, Science {\bf 368}, 6487, 173 (2020)
    
    
    \bibitem{FurtherIntriguingProperties} E. Bocquillon, V. Freulon,  F.D. Parmentier, J.-M. Berroir, B. Pla\c cais, C. Wahl, J. Rech, T. Jonckheere, T. Martin, C. Grenier, D. Ferraro, P. Degiovanni and G. F\`eve, , {\em Electron quantum optics in ballistic chiral conductors} Annalen der Physik, {\bf 526}, 1-30 (2014).
    
    \bibitem{FurtherIntriguingProperties2}
    F. Ronetti, L. Vannucci, D. Ferraro, T. Jonckheere, J. Rech, T. Martin, and M. Sassetti, {\em  Crystallization of levitons in the fractional quantum Hall regime}, Phys. Rev. B 98, 075401 (2018).



    \bibitem{Wen_topo} X. G. Wen, Adv. Phys. {\bf 44}, 405-473 (1995)
    \bibitem{Wen2} X. G. Wen, Int. J. Mod. Phys. B, {\bf 6}, 10, 1711-1762 (1992)
    \bibitem{Chang} A. M. Chang, Rev. Mod. Phys, {\bf 75}, 1449 (2003)
    \bibitem{ourepl} A. Nardin, I. Carusotto, EPL {\bf 132}, 10002 (2020)
    
    \bibitem{Yusa} A. Kamiyama, M. Matsuura, J. N. Moore, T. Mano, N. Shibata, G. Yusa, Phys. Rev. R. 4, L012040 (2022)
    
    \bibitem{Ashoori} R. C. Ashoori, H. L. Stormer, L. N. Pfeiffer, K. W. Baldwin, K. West ``Edge magnetoplasmons in the time domain", Phys. Rev. B 45, 3894 (1992)
    
    \bibitem{laughlin} R. B. Laughlin, Phys. Rev. Lett. {\bf 50}, 18, 1395 (1983)
    
    
    \bibitem{Wavefunctionology} S. H. Simon, ``Wavefunctionology", from ``Fractional Quantum Hall Effects: New Developments" edited by B. I. Halperin and J. K. Jain (2020). https://doi.org/10.1142/11751
        
    \bibitem{wilkingunn} N. K. Wilkin, J. M. F. Gunn, R. A. Smith, Phys. Rev. Lett. {\bf 80}, 2265 (1998)
    
    \bibitem{elia} E. Macaluso and I. Carusotto, {\em Hard-wall confinement of a fractional quantum Hall liquid}, Phys. Rev. A {\bf 96}, 043607 (2017).
    
    \bibitem{HaldanePseudoPotentials} F. D. M. Haldane, Phys. Rev. Lett. {\bf 51}, 7, 605 (1983).

    \bibitem{Trugman} S. A. Trugman, S. Kivelson, Phys. Rev. B, {\bf 31}, 8, 5280 (1985)   
    
    \bibitem{SimonRezayiCooper} S. H. Simon, E. H. Rezayi, N. R. Cooper, Phys. Rev. B {\bf 75}, 195306 (2007)
    
    \bibitem{Caz} M. A. Cazalilla, Phys. Rev. A {\bf 67}, 063613 (2003)

    \bibitem{RegnaultJolicoeur} N. Regnault, Th. Jolicoeur, Phys. Rev. B, {\bf 69} 235309 (2004)

    \bibitem{Stone} M. Stone, Phys. Rev. B {\bf 42}, 8399 (1990)

    \bibitem{ChamonWen} C. de C. Chamon, X. G. Wen, Phys. Rev. B {\bf 49} 8227 (1994)

    \bibitem{cooper_wedding_cakes} N. R. Cooper, 
    F. J. M. van Lankvelt, J. W. Reijnders, K. Schoutens, Phys. Rev. A, {\bf 72} 063622 (2005)
    
     \bibitem{MC_Jain0} J. K. Jain, R. K. Kamilla, ``Quantitative study of large composite-fermion systems", Phys. Rev. B {\bf 55}, R4895 (1997)
        
    \bibitem{MC_Jain1} C. C. Chang, N. Regnault, T. Jolicoeur, J. K. Jain, ``Composite fermionization of bosons in rapidly rotating atomic traps``, Phys. Rev. A {\bf 72}, 013611 (2005)
    
    \bibitem{MC_Jain2} S. Jolad and J. K. Jain, ``Testing the Topological Nature of the Fractional Quantum Hall Edge", Phys. Rev. Lett. 102, 116801 (2009)
    
    \bibitem{MC_Jain3} S. Jolad, D. Sen, J. K. Jain, ``Fractional quantum Hall edge: Effect of nonlinear dispersion and edge roton", Phys. Rev. B {\bf 82}, 075315 (2010)
    
    \bibitem{MC_Jain4} S. Jolad, C. C. Chang, J. K. Jain. ``Electron operator at the edge of the $1/3$ fractional quantum Hall liquid", Phys. Rev. B {\bf 75}, 165306 (2007)
     
    \bibitem{metropolis} N. Metropolis, A. W. Rosenbluth, M. N. Rosenbluth, A.
H. Teller, E. Teller, ``Equation of State Calculations
by Fast Computing Machines", J. Chem. Phys. 21, 1087
(1953)
     
    \bibitem{hastings} W. K. Hastings, ``Monte Carlo sampling methods using Markov chains and their applications", Biometrika 57 (1) 97-109 (1970)
     
    \bibitem{SandroBook} L. Pitaevskii, S. Stringari, ``Bose-Einstein Condensation and Superfluidity", Oxford Science Publications, Int. Series of Monographs on Physics. DOI:10.1093/acprof:oso/9780198758884.001.0001
    

    \bibitem{Imambekov} A. Imambekov, T. L. Schmidt, L. I. Glazman, Rev. Mod. Phys. {\bf 84} 1253 (2012).
    
    \bibitem{Glazman} A. Imambekov, L. I. Glazman, Science {\bf 323}, 5911, 228-231 (2009).
    
    \bibitem{Pustilnik} M. Pustilnik, M. Khodas, A. Kamenev, L. I. Glazman, Phys. Rev. Lett. {\bf 96}, 196405 (2006).
    
    \bibitem{Price} T. Price, A. Lamacraft, Phys. Rev. B {\bf 90}, 241415(R), (2014).
    
    \bibitem{NardinPRA} A. Nardin, I. Carusotto, in preparation (2022).
     
    \bibitem{Wiegman} P. Wiegmann, Phys. Rev. Lett. {\bf 108}, 206810 (2012).
     
    \bibitem{Wiegman2} E. Bettelheim, A. G. Abanov, P. Wiegmann, Phys. Rev. Lett. {\bf 97}, 246401 (2006).
     
    \bibitem{FernSimon} R. Fern, R. Bondesan, S. H. Simon, Phys. Rev. B, {\bf 98}, 15, 155321 (2018).
    
    \bibitem{Avron} J. E. Avron, R. Seiler, P. G. Zograf ``Viscosity of Quantum Hall Fluids", Phys. Rev. Lett. 75, 697 (1995)

    \bibitem{GMP} S. M. Girvin, A. H. MacDonald, P. M. Platzman, Phys. Rev. B {\bf 33} (4) 2481 (1986)
    
    \bibitem{KdV} O. Darrigol, {\em Worlds of flow: A history of hydrodynamics from Bernouilli to Prandtl} (Oxford University Press, Oxford, 2009).
    
    \bibitem{KdV2} N. J. Zabuski and M. D. Kruskal, Phys. Rev. Lett. {\bf 15}, 240 (1965).
    
    \bibitem{Bretin} V. Bretin, S. Stock, Y. Seurin, J. Dalibard  ``Fast Rotation of a Bose-Einstein Condensate",
Phys. Rev. Lett. 92, 050403 (2004)
    
    \bibitem{Schweikhard}
V. Schweikhard, I. Coddington, P. Engels, V. P. Mogendorff, E. A. Cornell ``Rapidly Rotating Bose-Einstein Condensates in and near the Lowest Landau Level",
Phys. Rev. Lett. 92, 040404 (2004)
    


    \bibitem{Andrade} B. Andrade, V. Kasper, M. Lewenstein, C. Weitenberg, T. Gra\ss{}, ``Preparation of the $1/2$ Laughlin state with atoms in a rotating trap", Phys. Rev. A 103, 063325 (2021)
    
    \bibitem{Zwierlein2} J. Fletcher, Am Shaffer, C. C. Wilson, P. B. Patel, Z. Yan, V. Crépel, B. Mukherjee, M. W. Zwierlein ``Geometric squeezing into the lowest Landau level", Science, {\bf 372}, 6548, 1318-1322 (2021) 
    
    \bibitem{Gaunt} A. L. Gaunt, T. F. Schmidutz, I. Gotlibovych, R. P. Smith, Z. Hadzibabic ``Bose-Einstein Condensation of Atoms in a Uniform Potential", Phys. Rev. Lett. 110, 200406 (2013)
    
    \bibitem{dalibard} F. Chevy, K. W. Madison, and J. Dalibard ``Measurement of the Angular Momentum of a Rotating Bose-Einstein Condensate", Phys. Rev. Lett. 85, 2223 (2000)
    
    \bibitem{ZenoMSc} Zeno Bacciconi, {\em Fractional quantum Hall edge dynamics from a quantum optics perspective}, MSc thesis at Trento-SISSA joint program in physics, available as arXiv:2111.05858.
    
    \end{thebibliography}
\end{document}